\documentclass[a4paper,11pt,leqno]{amsart}

\usepackage{amsmath}
\usepackage{amsthm}
\usepackage{amsfonts}
\usepackage{amssymb}

\usepackage{tikz}
\usetikzlibrary{3d, calc,shadings}
\usepackage{enumerate}
\usepackage{colonequals}
\usepackage{booktabs}
\usepackage{url}
\usepackage{todo}
\usepackage{subcaption}
\usepackage{anyfontsize} 
\usepackage{pdflscape}
\usepackage{algorithm}
\usepackage{algpseudocode}
\usepackage{graphicx}
\usepackage{adjustbox}

\newcommand{\eps}{\varepsilon}
\newcommand{\N}{\mathbb N}

\newcommand{\R}{\mathbb R}

\newtheorem{theorem}{Theorem}[section]
\newtheorem*{theorem*}{Theorem}{\bf}{\it}

\newtheorem*{proposition*}{Proposition}{\bf}{\it}

\newtheorem*{lemma*}{Lemma}{\bf}{\it}

\theoremstyle{definition}
\newtheorem{definition}[theorem]{Definition}
\newtheorem*{definition*}{Definition}
\theoremstyle{remark}

\numberwithin{equation}{section}

\numberwithin{equation}{section}

\title[Geometry of Finnish roads]{Observing rurality of a geographical area from road graph geometry -- a qualitative study}
\author{$\R$ami Luisto}
\address{
Faculty of Information Technology, P.O. Box 35 (Mattilanniemi 2), FI-40014 University of Jyv\"askyl\"a, Finland
\and
Digital Workforce Services
Mechelininkatu 1 a,
00180 Helsinki, Finland
\and
Department of Mathematics and Statistics, P.O. Box 68 (Pietari Kalmin
katu 5), FI-00014 University of Helsinki, Finland
\and
Helsinki Institute of Urban and Regional Studies (Urbaria),
P.O. Box 3
(Fabianinkatu 33)
00014 University of Helsinki}
\email{rami.luisto@gmail.com}

\subjclass[2010]{51F99, 00A05, 90B20}
\date{\today}

\begin{document}

\maketitle


\begin{abstract}
    The driving motivation behind this paper is a personal observation on Finnish roads and their geometry, which we try to visualize in Figure \ref{fig:road_geometry_comparison}.
\end{abstract}

\begin{figure}[h!]
    \centering
    \begin{subfigure}{0.45\textwidth}
        \centering
        \adjustbox{trim={.08\width} {.12\height} {.04\width} {.14\height},clip}%
            {\includegraphics[width=\textwidth]{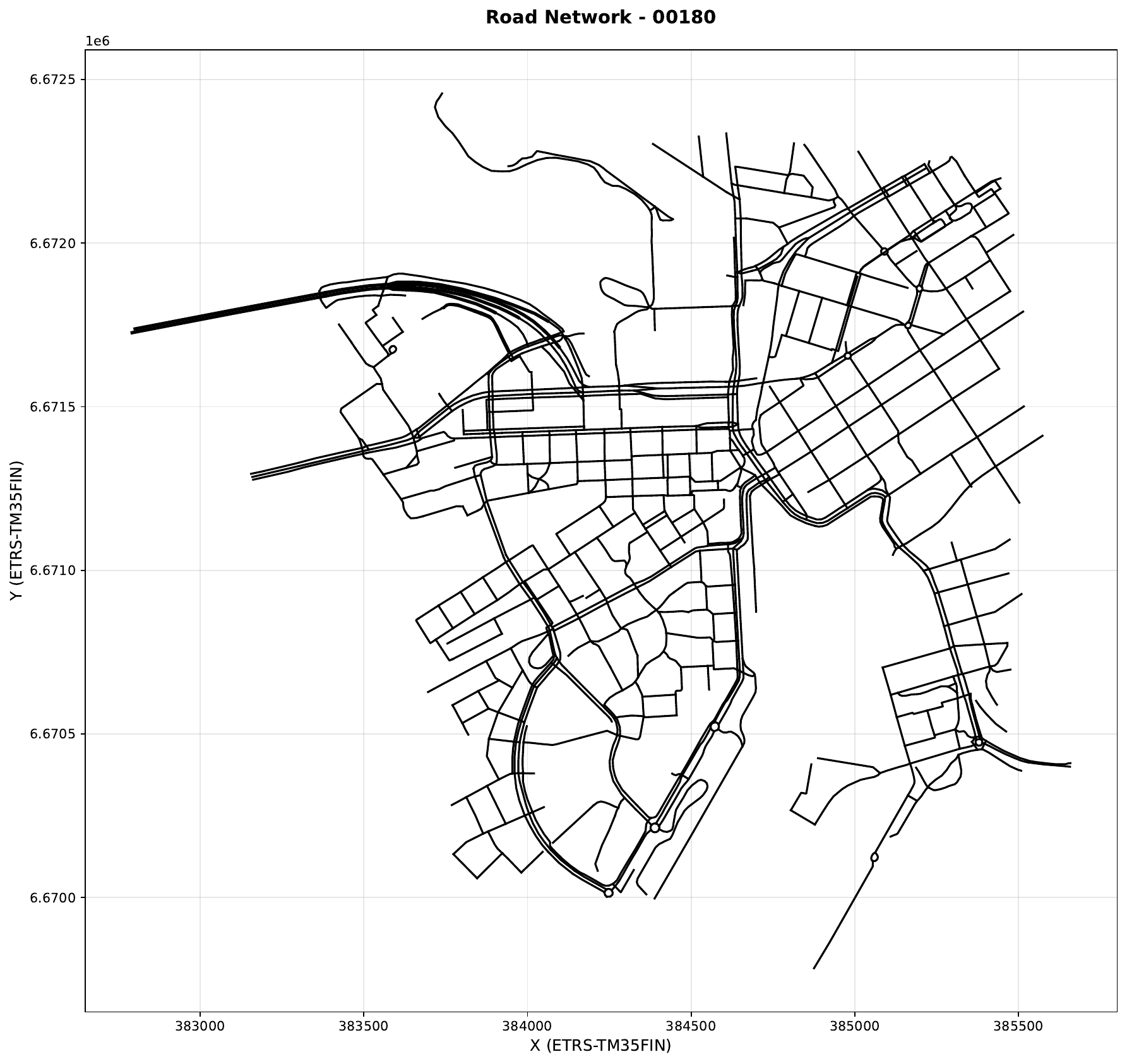}}
        \caption{Kamppi-Ruoholahti}
    \end{subfigure}
    \hfill
    \begin{subfigure}{0.45\textwidth}
        \centering
        \adjustbox{trim={.08\width} {.08\height} {.04\width} {.07\height},clip}%
            {\includegraphics[width=\textwidth]{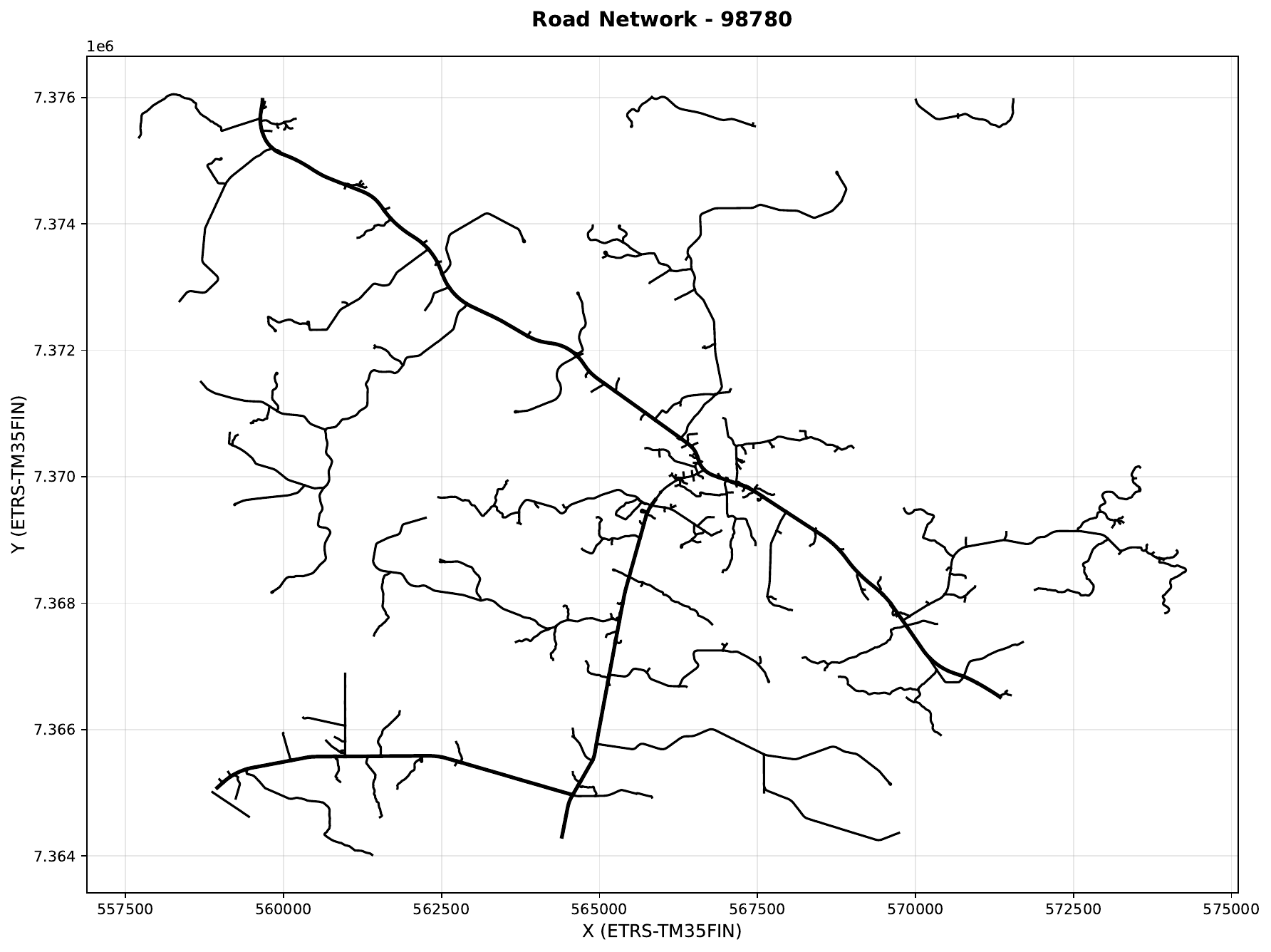}}
        \caption{Maaninkavaara}
    \end{subfigure}
    
    \vspace{0.5cm}
    
    \begin{subfigure}{0.45\textwidth}
        \centering
        \adjustbox{trim={.10\width} {.10\height} {.05\width} {.10\height},clip}%
            {\includegraphics[width=\textwidth]{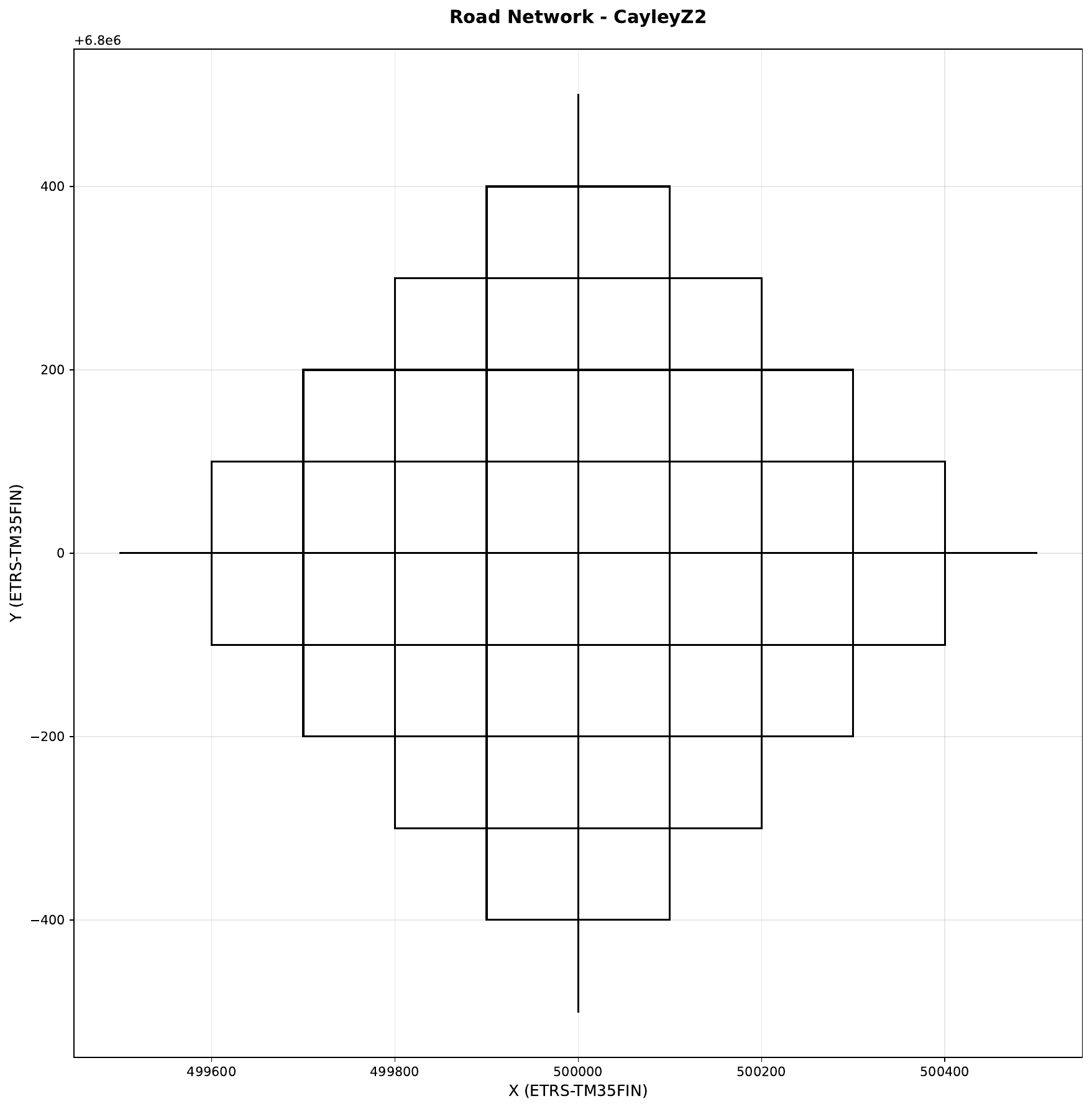}}
        \caption{Abelian Group}
    \end{subfigure}
    \hfill
    \begin{subfigure}{0.45\textwidth}
        \centering
        \adjustbox{trim={.10\width} {.08\height} {.05\width} {.08\height},clip}%
            {\includegraphics[width=\textwidth]{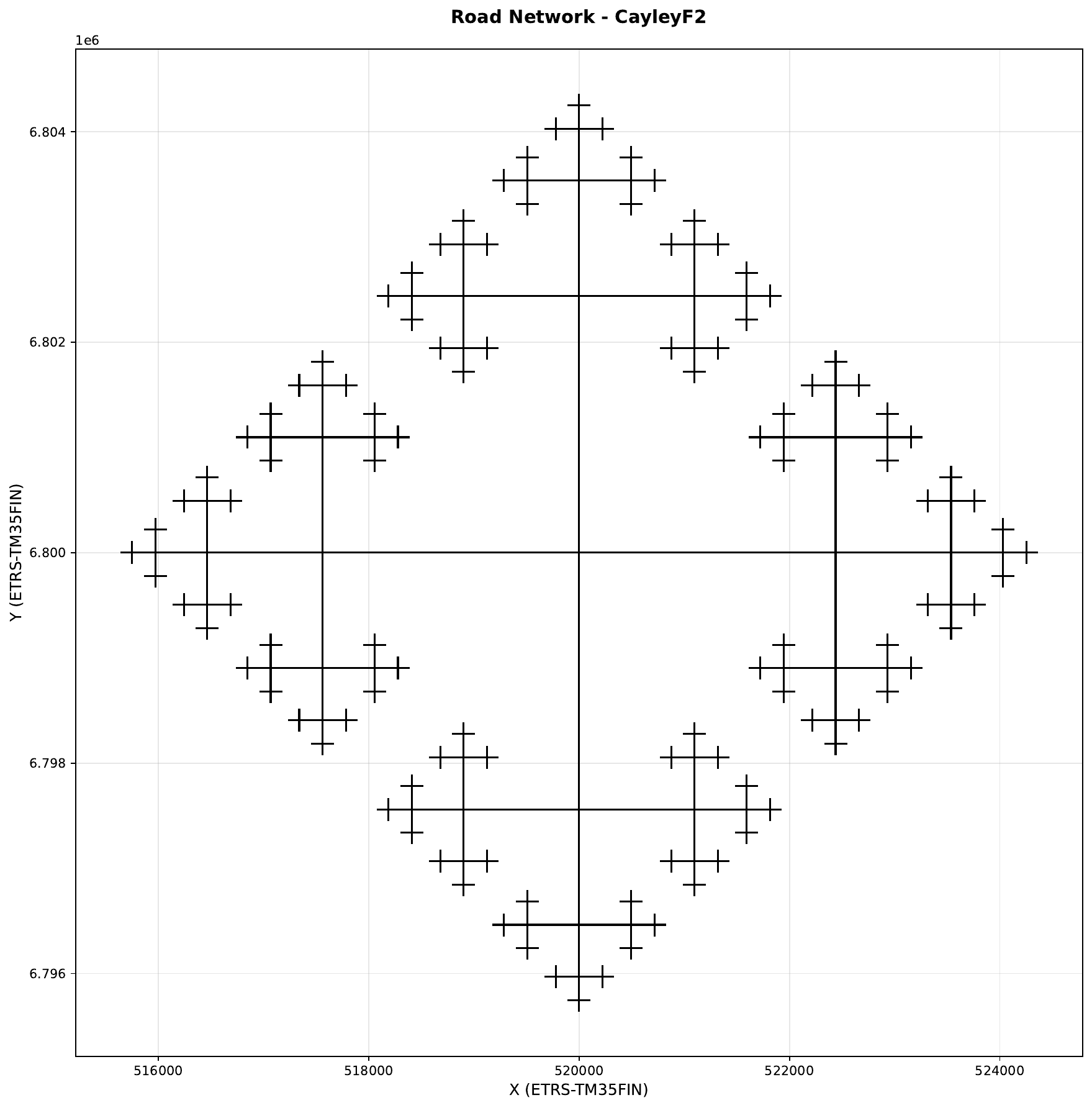}}
        \caption{Free (hyperbolic) Group}
    \end{subfigure}
    
    \caption{Various graph geometries found in road systems and groups.}
    \label{fig:road_geometry_comparison}
\end{figure}

\section{Introduction}
\label{sec:Intro}

The purpose of this paper is to describe how the networks of roads in rural areas seem to have a more hyperbolic geometry than the road networks in urban areas. We provide statistical data on various computational ``hyperbolicity measures'' of such road networks, but we emphasize that our aim is to \emph{describe} and \emph{observe} the phenomenon qualitatively rather than to give quantitative estimates. The core idea is shown in Figure \ref{fig:road_geometry_comparison} where the top row shows road networks in downtown Helsinki and rural Lapland while the bottom row shows parts of the Cayley graphs of the Abelian and free groups of two generators -- the similarity of the graph structure within both columns is what we hope to measure here.

For the mathematician reader of this paper we wish to emphasize that we are not aiming for proofs. We have furthermore taken some liberties in adapting mathematically exact concepts to the setting of numerical real-world data. We might describe the work here not as \emph{mathematics} but an analysis \emph{inspired by} notions of metric geometry and hyperbolic spaces.

The structure of this paper is as follows. We will begin by giving a brief general level introduction to some concepts of hyperbolicity in metric spaces in Section \ref{sec:GeneralHyperbolicity}. That section will contain both a more heuristical approach and some more techincal notions. After that we'll describe in Section \ref{sec:DataAndMethods} the data and methods we have for this project. (The source code behind our systems is available at \url{https://github.com/ramiluisto/RoadGeometry}.)
We then describe some of our statistical findings in Section \ref{sec:results}. Finally we conclude the paper with some final thoughts in Section \ref{sec:conclusions}.

\section{A few notes about hyperbolicity in metric spaces}
\label{sec:GeneralHyperbolicity}

In the field of metric geometry, there are many notions that correspond to curvature and hyperbolicity in various ways. After studying several of them, one learns a sort of holistic mental concept of these geometrical notions. It is out of the scope of this paper to try to introduce to the reader the full scope of these ideas, but in this section our aim is to give some general overview on how the notions of geodesics, triangles, commutativity and volume growth estimates fit together in the realm of hyperbolic geometry. For a more thorough general audience introduction we refer to the excellent \cite{weeks2001shape}.

\subsection{Some heuristical observations on geometry and curvature}

In the classical 2-dimensional geometry there are three ``archetypical'' classes of curvature: \emph{hyperbolic} (negative curvature), \emph{flat} (zero curvature) and \emph{spherical} (positive curvature). Flat geometry is the geometry of the standard Euclidean plane studied since the ancient Greeks. Spherical geometry is what we see in the surface of a sphere, and hyperbolic geometry is often depicted with a sort of saddle-looking surface. The hyperbolic geometry tends to be the least familar of these three as there are very few natural examples of hyperbolic objects in everyday life.

There are many ways to approach and formalize these classes of geometry, but in this section our baseline is based on volume\footnote{We'll opt for the term \emph{volume} and \emph{balls} here instead of \emph{area} and \emph{disks}. We reserve the term \emph{area} only for shapes in the Euclidean space and use \emph{volume} for the more general notion applicable in higher dimensions and non-Euclidean spaces.} growth:
\begin{itemize}
    \item In the flat geometry the volume of balls of radius $r$ grow as $\mathcal{O}(r^2)$ (e.g.\ in the plane the area of the disk is $\pi r^2$). This flat geometry is the one to which we compare other geometries.
    \item In spherical geometry the volume of balls grow \emph{slower} than in the plane. I.e.\ they ``have less volume in them''.
    \item In hyperbolic geometry the volume of balls grow \emph{faster} than in the plane. I.e.\ they ``have more volume in them''.
\end{itemize}

\begin{figure}[h!]
    \centering
    \begin{subfigure}{0.3\textwidth}
        \centering
        \includegraphics[width=\textwidth]{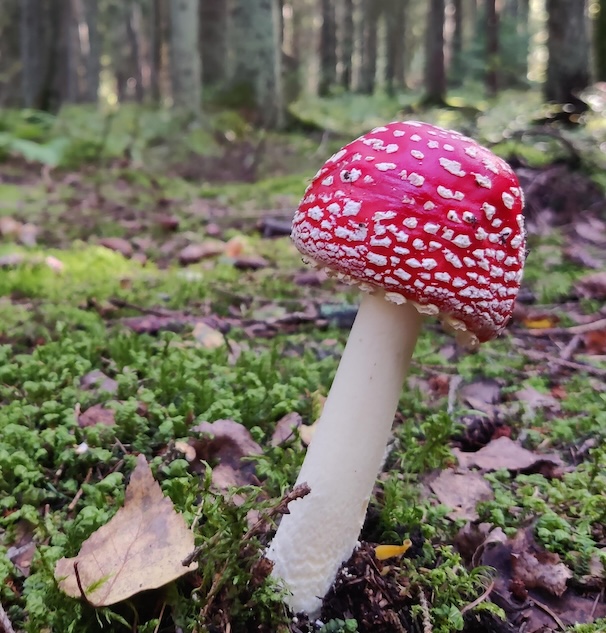}
    \end{subfigure}
    \hfill
    \begin{subfigure}{0.3\textwidth}
        \centering
        \includegraphics[width=\textwidth]{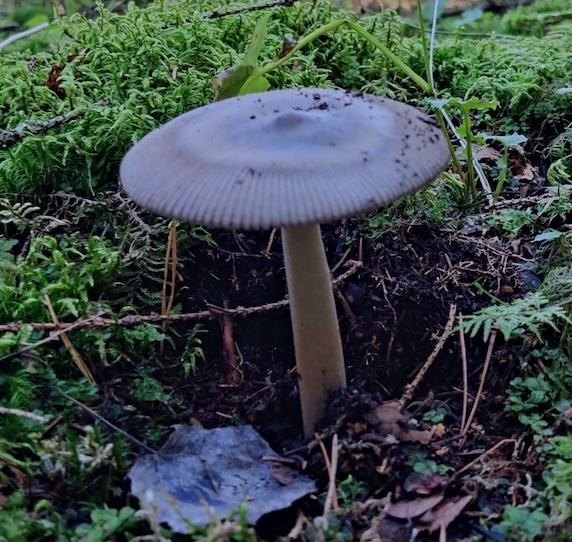}
    \end{subfigure}
    \hfill
    \begin{subfigure}{0.3\textwidth}
        \centering
        \includegraphics[width=\textwidth]{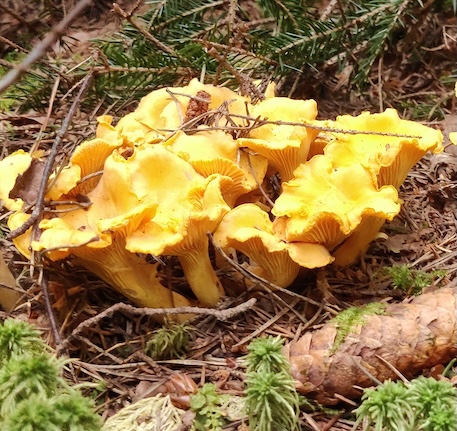}
    \end{subfigure}

    \caption{Examples of mushroom caps illustrating different geometries: spherical (fly agaric), flat (amethyst deceiver) and hyperbolic (chantarelle).}
    \label{fig:mushroom_geometry}
\end{figure}

For us the standard visualization for these differences in geometry is found from mushrooms; see Figure \ref{fig:mushroom_geometry}. Imagine a mushroom that is growing more and more cap. Let's imagine that it is growing it like a knitter knits a hat; adding more rings to the outer edge. It has to grow the cap with a very specific speed to end up with a completely flat cap. If the added rings have less volume than that, the mushroom will ``pull'' its geometry more spherical. If it grows more, the cap will turn all wavy. This approach provides excellent ways of demonstrating and visualizing some core concepts of hyperbolicity through knitting, see e.g.\ \cite[Chapter: Crocheting mathematics]{ShapesInAction} or \cite{taimina2018crocheting}.

Already in this approach we see reflected some more general trends. Spherical geometry tends to curl up on itself\footnote{See e.g.\ Myer's theorem \cite{myers1941riemannian}.}, while hyperbolic geometries can be hard to visualize as they don't fit in Euclidean spaces very well. We again refer to Jeffrey Week's \emph{The Shape of Space} (\cite{weeks2001shape}) for a more detailed but still general introduction which includes good ways visualize and understand hyperbolic spaces.

The volume argument is something that, surprisingly, ties into the algebraic concept of ``commutativity''. In algebra, commutativity refers to the idea that the order of operation(s) does not matter. E.g.\ the addition and multiplication of real numbers are commutative as $a+b = b+a$ and $ab = ba$. Many other natural operations are not; if you have e.g. a book in front of you, then rotating it 90 degrees clockwise and flipping horizontally produces a different end result than first flipping it horizontally and then rotating it 90 degrees clockwise.

In these abstract mathematical object called groups we can, in certain situations\footnote{Typically in finitely generated groups, see e.g.\ \cite{mann2011groups}.}, define a concept called \emph{growth rate} which measures how many different group elements can be expressed by using at most $n$ elements of a finite spanning set. Suppose we have only two elements in the spanning group, $a$ and $b$. Then we can start to construct more elements as e.g.\ $aa$, $ab$, $ba$, $bb$ and so forth. Now the crucial question is about how many of such elements are actually the same element written differently, like $2 \cdot 3 = 3 \cdot 2$. What we note immediately is that if $ab=ba$, i.e.\ if these elements commute, then there are less new elements created! This means that with less commutability the growth rate of the group is bigger, i.e.\ there is more volume in it, i.e.\ it is more hyperbolic. (We note that there is \emph{a lot} more nuance to the concept of commutativity and growth rate; see e.g.\ Gromov's theorem \cite{gromov1981groups} which states that a finitely generated group has polynomial growth rate if and only if it is virtually nilpotent.) In any case, the two Cayley graphs in the bottowm row of Figure \ref{fig:road_geometry_comparison} demonstrate the two extremal cases; an Abelian group and a free group. The abelian group has a so called \emph{polynomial growth rate}\footnote{We will not go to details here, see again e.g.\ \cite{mann2011groups}, but in each graph if you fix a starting point and count how many elements can be reached after $n$ steps in the graph, you will find $\mathcal{O}(n^2)$ elements in the Cayley graph on the left and $\mathcal{O}(2^n)$ elements on the Cayley graph on the right.} while the free group has an \emph{exponential} one, but these should be considered as the ends of a spectrum. For our purposes, the crucial note is that reduced commutativity correlates with increased volume growth.

\begin{figure}[htbp]
    \centering
    \includegraphics[width=0.8\textwidth]{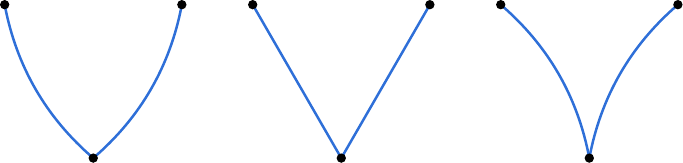}
    \caption{Basic images of geodesics in spherical, flat and hyperbolic spaces.}
    \label{fig:straight_lines_side_by_side}
\end{figure}

Another geometric approach here is to look at the idea that in a hyperbolic space more volume means, in a sense, more \emph{directions} to go to. The classical image to draw here is that if you start from a point and look at two ``straight lines'' extending from it, then in a hyperbolic group they veer off faster than in a Euclidean space, while in a spherical space they turn towards each other, see Figure \ref{fig:straight_lines_side_by_side} for the standard visualization. This veering off happens, in a sense, since there is more area between the two lines. Once again, the spherical case is easier to visualize as we can imagine two geodesics on the surface of a sphere, or a cap of a spherical mushroom, converging towards each other. This effect has a strong influence on the geometry of triangles, which will prove to be important for our numerical methods.

Besides volumes, the direction based point of view to hyperbolicity also ties to the ideas of commutativity. From our road-centric view, imagine for a moment the route options of a big city center like Manhattan. In most cases turning right and then left equals going straight and then right since it simply brings you around the block. In a more rural system, if you miss a turn you will quite likely have to back up all the way to where you made the mistake. This lack of commutativity thus produces more directions and, in a sense, more ``volume'' to the road system.

And this brings us back to the core idea of this paper. From a holistic point of view of typical hyperbolic behaviours, the road systems of rural areas do start to seem more hyperbolic than flat. We'll next turn to some more technical notions on how we could measure this in a more quantitative way.

\subsection{Measuring hyperbolicity in a metric space}
\label{sec:MeasuringHyperbolicity}

In studying the road systems, a natural approach will be to consider the road systems as metric spaces where the metric is defined e.g.\ by distance or travel time between two points. There is ample literature on curvature an hyperbolicity in metric spaces, and we'll simply mention our personal favorites: \cite{burago2001course, gromov2007metric,vaisala2005gromov}. Instead of trying to give a proper mathematical introduction, we'll simply describe the methods most suitable for us, and then on how to adapt them into our situation. We wish to emphasize that none of the concepts discussed here are novel, and can be found e.g.\ from the sources just mentioned.

Our methods of measuring the ``level of hyperbolicity'' is based on studying triangles in a metric space. To discuss these we'll set up some basic terminology. For readers unfamiliar with mathematics of metric spaces we recommend them to skim (or skip) the rest of the section and focus on the concrete measures we define in Section \ref{sec:DataAndMethods}.

\begin{definition}
    Let $(X,d)$ be a metric space. A \emph{path} (in $X$) is a continuous mapping $\alpha \colon [0,1] \to X$. If $\alpha(0) = x$ and $\alpha(1) = y$, then we say that $\alpha$ is \emph{a path connecting $x$ to $y$} and denote $\alpha \colon x \curvearrowright y$.
\end{definition}

For us a crucial notion in regards to paths will be their length. Note that we are here working with more abstract distances and lengths than just the Euclidean standard ones. In particular, for use the ``length'' of a path can be either its length in meters or the travel time it would take to drive through it with maximum allowed speed.
\begin{definition}
    The \emph{length} of a path $\alpha \colon [0,1] \to X$ is set to be
    \begin{align*}
        \ell(\alpha) 
        \colonequals \sup \left\{
        \sum_{j=1}^n d(\alpha(a_{j-1}), \alpha(a_j)) 
        \mid 
        a_0 < \cdots < a_n, n \in \N
        \right\} \in \R \cup \{ \infty \}.
    \end{align*}

    Note that the length of a path can be infinite. We call a metric space \emph{rectifiably connected} if any two points can be connected with a path of finite length. For rectifiably connected metric spaces we can define the \emph{path metric} $d_\ell$ by setting
    \begin{align*}
        d_\ell(x,y) 
        \colonequals \inf \left\{ 
            \ell(\alpha) \mid \alpha \colon x \curvearrowright y
        \right\}.
    \end{align*}
\end{definition}

\begin{figure}[htbp]
    \centering
    \includegraphics[width=0.8\textwidth]{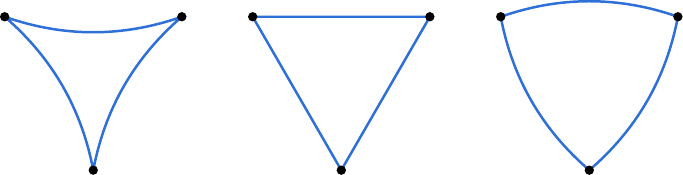}
    \caption{Basic images of triangles in hyperbolic, flat and spherical spaces.}
    \label{fig:triangles_side_by_side}
\end{figure}

With this terminology we can finally define a triangle, which for us will be a triplet of points where the sides are not (necessarily) straight lines but any shortest path between them.
\begin{definition}
    Let $(X,d)$ be a metric space. We call a path $\alpha \colon x \curvearrowright y$ a \emph{geodesic} if $\ell(\alpha) = d(x,y)$. Note that in general geodesics do not need to be unique.

    A \emph{triangle} in a metric space is a triplet of points $a,b,c \in X$ together with geodesics $\alpha \colon a \curvearrowright b$, $\beta \colon b \curvearrowright c$ and $\gamma \colon c \curvearrowright a$.
\end{definition}
In Figure \ref{fig:triangles_side_by_side} we show the typical examples of these metric triangles in the three typical geometries.

With this set up, we can now start to estimate levels of hyperbolicity in spaces based on triangles. Since our objects of study are finite road networks, it is challenging to actually apply most of the definitions of a hyperbolic metric space here. But we will list a few techniques to estimate the hyperbolicity of a single triangle, and later on we'll build on these by sampling such triangles from the road networks.

\subsection{Numerical measurements}

All of our measures in this paper will focus on differentiating more hyperbolic triangles from flatter triangles. Or rather, trying to measure the ``level of hyperbolicity'' of a triangle. We will often aim to make things more scale invariant by applying the classical approach of \emph{comparison triangles}\footnote{Comparison triangle approaches are especially used in the study of the so called $\operatorname{CAT}(\kappa)$ spaces, see e.g.\ \cite[Chapter 4]{burago2001course} or \cite{gromov2007metric}.}. By this we mean that given a triplet of points $(a,b,c)$ in a metric space, we can always find a triplet of points $(a', b', c')$ in the Euclidean plane with the same pairwise distances. Then for any measure we build for the metric triangle we can replicate for such a comparison triangle and compare them e.g.\ through ratios. For a more in depth introduction to this, see again \cite[Chapter 4]{burago2001course}.

\begin{figure}[htbp]
    \centering
    \begin{subfigure}{0.45\textwidth}
        \centering
        \includegraphics[width=\textwidth]{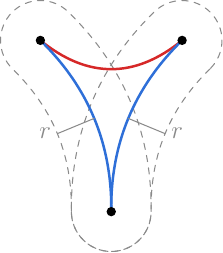}
        \caption{Hyperbolic triangle}
    \end{subfigure}
    \hfill
    \begin{subfigure}{0.45\textwidth}
        \centering
        \includegraphics[width=\textwidth]{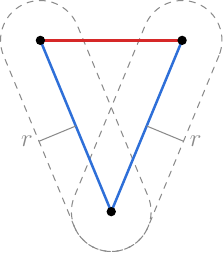}
        \caption{Flat triangle}
    \end{subfigure}
    \caption{A comparison of the Hausdorff distance of a given edge of a triangle to the two other edges.}
    \label{fig:hausdorff_triangles}
\end{figure}

The first classical approach is described in Figure \ref{fig:hausdorff_triangles}. Here the idea is that unlike in a flat triangle, in a hyperbolic triangle any side is quite close to the two other sides\footnote{This is usually described by saying that hyperbolic triangles are \emph{$\delta$-slim} and a metric space where all geodesic triangles are $\delta$-slim are called \emph{Gromov $\delta$-hyperbolic}, see \cite{burago2001course}.}. We can measure this by asking what is smallest radius $r$ such that any given side is contained in the $r$-neighbourhood of the two other sides. In a standard Euclidean triangle this is just the maximum height of the triangle. 

\begin{figure}[htbp]
    \centering
    \begin{subfigure}{0.45\textwidth}
        \centering
        \includegraphics[width=0.7\textwidth]{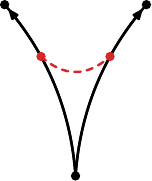}
    \end{subfigure}
    \hfill
    \begin{subfigure}{0.45\textwidth}
        \centering
        \includegraphics[width=\textwidth]{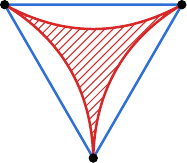}
    \end{subfigure}
    \caption{Veer-off speed (left) and area comparison (right) of a hyperbolic triangle.}
    \label{fig:triangle_speed_area}
\end{figure}

Another approach is to study how fast the two sides of a triangle start to ``veer off'' in a hyperbolic triangle, see left side of Figure \ref{fig:triangle_speed_area}. Here the idea is to imagine two sides of a triangle ``eminating'' from the joint starting point and measure how fast two points traveling ``at unit speed'' gain distance to each other. We did not end up using this approach in this paper, but mention it as it would be an interesting further study target.

We can also study the area of a triangle, especially compared to a comparison triangle; see the right side of Figure \ref{fig:triangle_speed_area}. This approach requires that we have some method of calculating the area of a metric triangle. In general this might not be possible, and often requires that we are studying a 2-manifold with a measure that fits in together with the distance\footnote{Technically, we would most likely want to have a 2-manifold which is e.g.\ a doubling path-metric measure space. With some extra caveats so that we can for a given metric triangle decide which part is inside or outside.}. In our case we will simply abuse this notion and mix up two domains; we'll draw the geodesics based on our road network metric space and then calculate the area as if the metric triangle was an Euclidean shape\footnote{Here in particular we note that our measure is no longer an \emph{intrinsic} measure of the road network but is dependent on how it is actually situated in the world. This is not inherently a bad thing, but should be noted.}.

Finally we can study the shape of objects through their \emph{isoperimetric inequality}. These inequalities in general refer to comparing the ratio between the length of the boundary of an object to its area. (Or the area of its boundary to its volume in three dimensions.) Here we face the same issues as with the previous area-comparison approach: we have to ``mix domains'' and measure the areas in the Euclidean plane where the roads are set\footnote{Technically they of course lie on the \emph{spherical} geometry of Earth, but for our purposes we work with flat maps of the road structures.}, which turns this measure non-intrinsic. There are several variations on how to exactly measure the ``isoperimetricity'' of a shape, but in this paper we'll use the Polsby-Popper test\footnote{Originally developed to measure the level of Gerrymandering in political districts, \cite{gerrymandering1991third}, which we feel fits our domain of geographical data.} which divides the area of a shape with its perimeter squared, scaled so that a circle would score $1.0$.

\section{Our roads and methods}
\label{sec:DataAndMethods}

\subsection{The data}
\label{sec:data}

Our road geometry analysis is based on the DIGIROAD K dataset \cite{FTIA_Digiroad_2025} by the Finnish Transport Infrastructure Agency who maintain several high quality open source datasets on Finnish roads (among other things).
The data in the DIGIROAD K material is divided to small road segments called ``links''. These have, among other data, their end points and speed limit recorded. They do not contain (to our knowledge) direct connection data on how they connect, but we've opted to connect any two link endpoints whose distance is less than one meter. This connection approach has most likely introduced some errors in our road network graphs, but in our statistical approach this should not be too detrimental. The dataset has also information on which road segments are one-way streets, but for simplicity we've omitted this information in our analysis and pretend that all streets are two-way.\footnote{Our analysis tools can take this directionality into account, but the results seemed to stay largely the same.}
We've further used the Statistics Finland Paavo dataset \cite{StatisticsFinland_Paavo_2025} which contains demographic information on the level of postal code areas toghether with their geometric boundaries.

Combining these two datasets we're then able to extract the road networks within a given postal code area. We use the population information and the geometric information to calculate an average population density for each postal code area. This population density is our simple proxy variable on the level of rurality that an area has. Though we note that the Paavo dataset divides the postal code areas to three classes of urban, semiurban and rural, and we also look at this datum a bit in the results section.

\subsection{The methods}
\label{sec:methods}

Our core idea is that we analyze areas one postal code at a time. Our analysis tools can handle larger regions, but as we've not optimized the system for speed or efficiency, larger than postal code areas require considerably more time to analyze, thus limiting large scale analysis. We'll discuss some of the limitations of this effect in Secion \ref{sec:conclusions}.

\begin{figure}
    \centering
    \adjustbox{trim=0 {.29\height} 0 0,clip}%
            {\includegraphics[width=\textwidth]{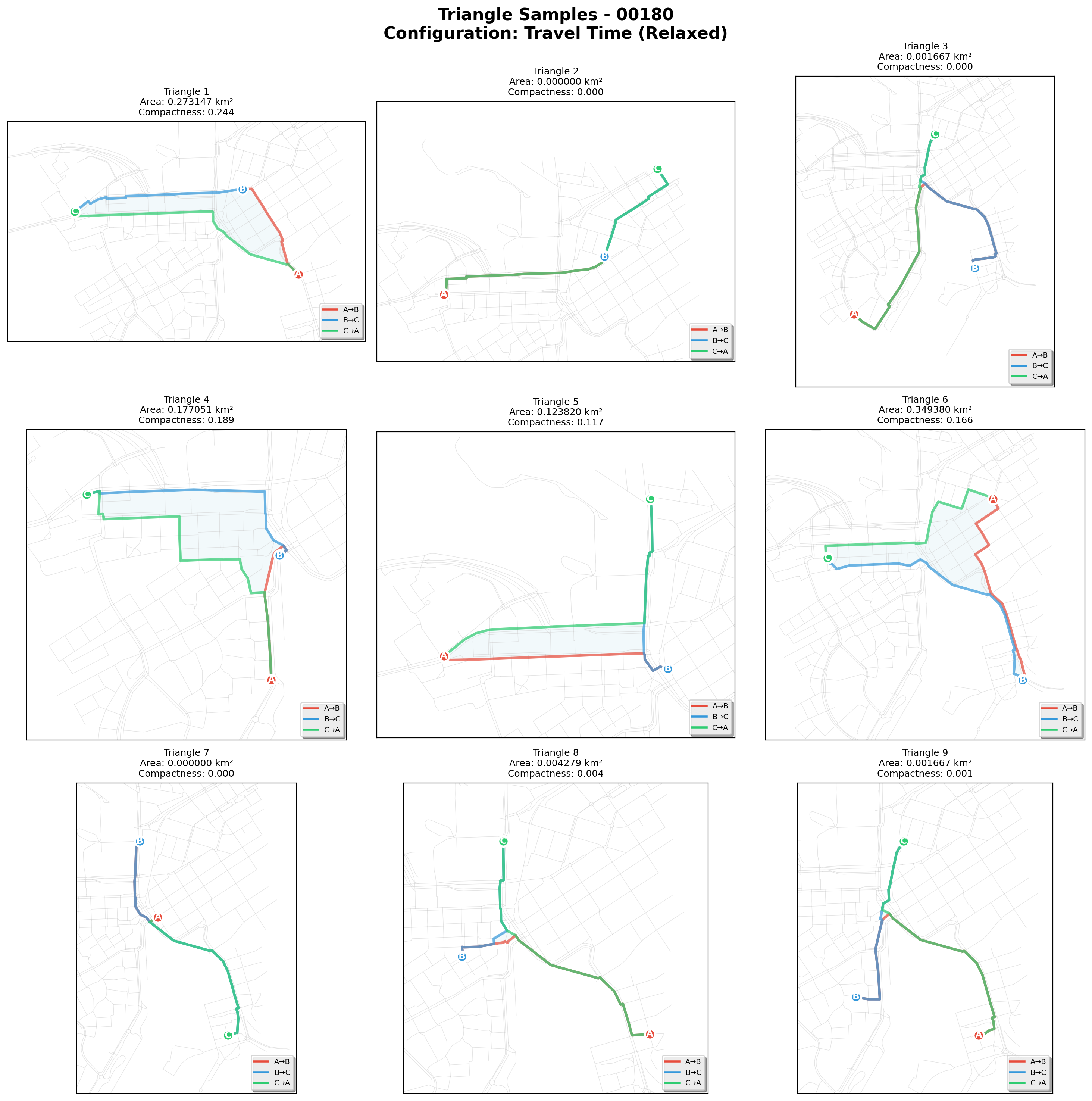}}
    \caption{A few randomly sampled travel-time geodesic triangles in the 00180 postal code area.}
    \label{fig:00180_example_triangles}
\end{figure}

From a given postal code region we sample triplets of points by sampling them from the bounding box of the region and then projecting them to the nearest road end point. For each such triplet we then find geodesics connecting each pair. We've run our various analyses with two kinds of geodesics: those of minimal length and those of minimal travel time. We'll focus mostly on the length geodesics here for the sake of compactness of exposition, the travel time results were somewhat similar; we'll show one comparison in Section \ref{sec:conclusions}. In Figure \ref{fig:00180_example_triangles} we show a few randomly sampled travel-time triangles in the urban 00180 postal code area.

For these triangles we calculate the following measures:
\begin{itemize}
    \item \textbf{Compactness.} This is the the isoperimetric Polsby-Popper test where we divide the area of the triangle with the perimeter length squared and scale it to give the value $1.0$ for circles.
    \item \textbf{Comparison Triangle Ratio.} Here we take the triangle and divide its area with the area of a comparison triangle, i.e.\ an Euclidean triangle whose vertices have the same pairwise distances as our hyperbolic triangle.
    \item \textbf{Relative Distortion.} In this method we find the smallest radius $r$ such that any side of the triangle is contained in the $r$-neighbourhood of the two other sides -- i.e.\ we find the $r$-slimness of the triangle. We then divide this number with the corresponding number we would have for an Euclidean triangle.
\end{itemize}

\section{Results}
\label{sec:results}

We'll go through our results in two stages. First we look at three specific areas, hand picked to correspond to urban, semiurban and rural areas. Then we'll look at some statistical results from a larger sampling of areas.

\subsection{Results in three representative areas}

\begin{figure}
    \centering
    \begin{subfigure}[t]{0.32\textwidth}
        \centering
        \includegraphics[width=\textwidth]{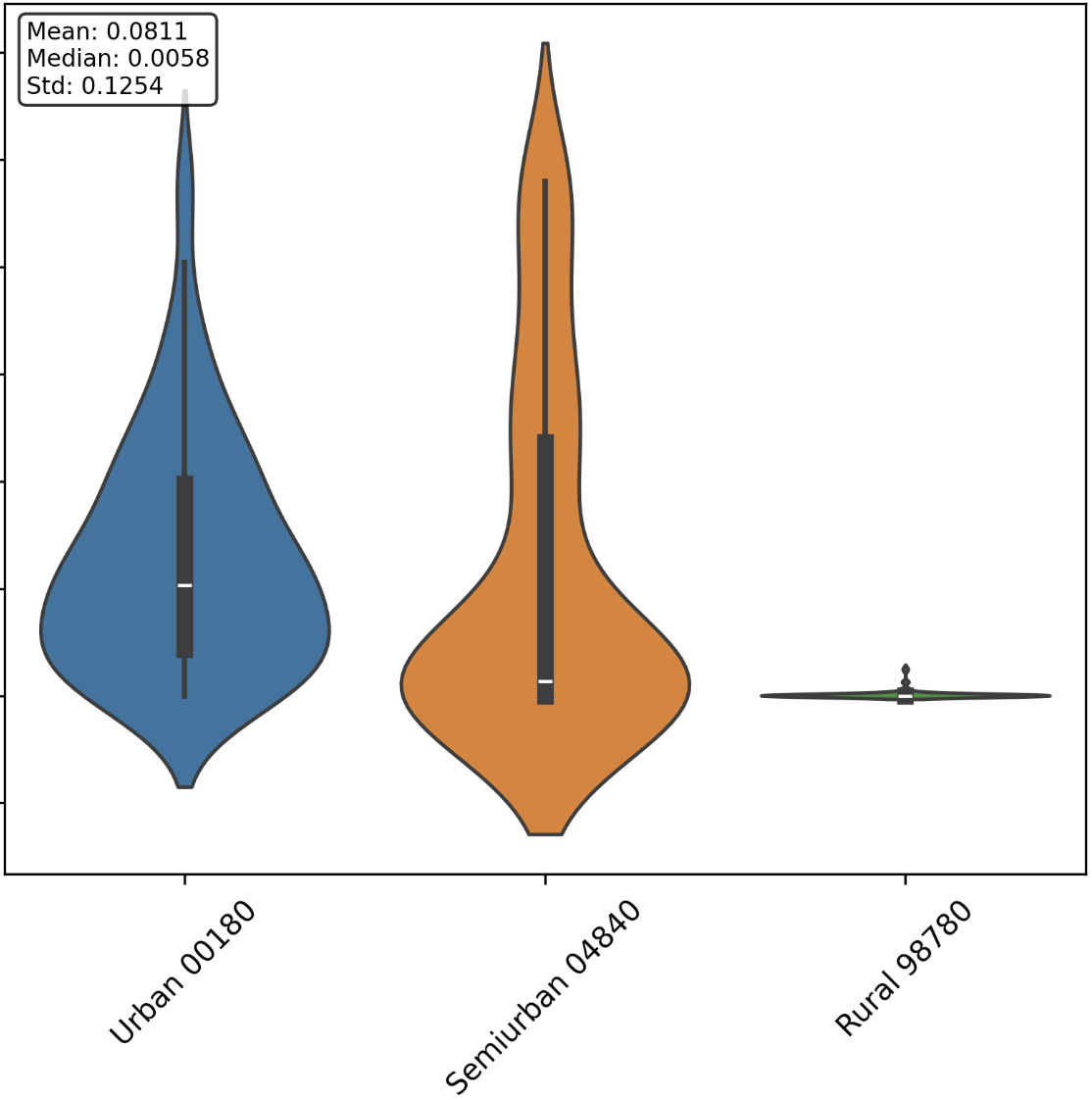}
        \caption{Compactness (Polsby-Popper)}
        \label{fig:compactness}
    \end{subfigure}
    \hfill
    \begin{subfigure}[t]{0.32\textwidth}
        \centering
        \includegraphics[width=\textwidth]{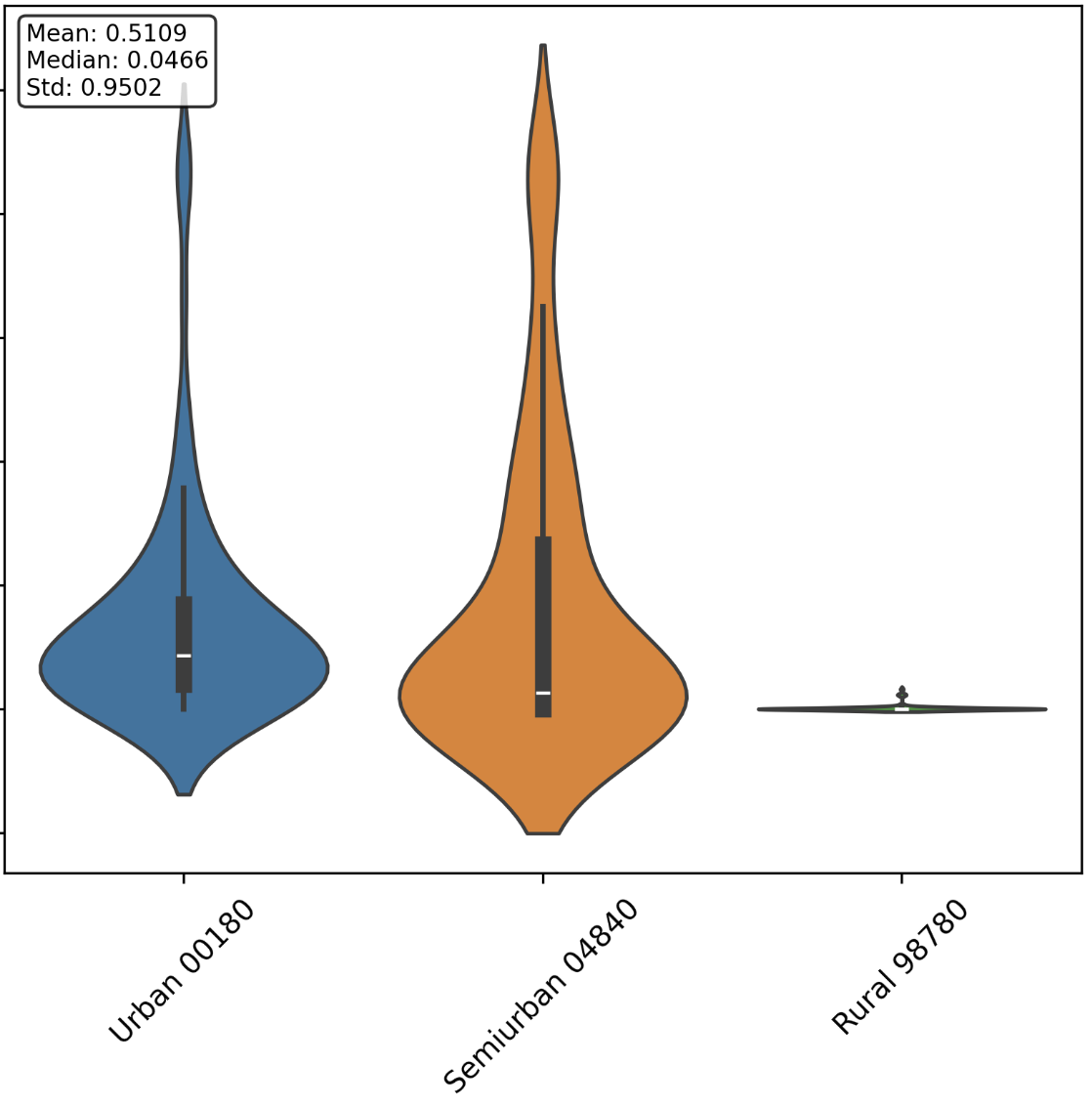}
        \caption{Comparison Triangle Ratio}
        \label{fig:triangle}
    \end{subfigure}
    \hfill
    \begin{subfigure}[t]{0.32\textwidth}
        \centering
        \includegraphics[width=\textwidth]{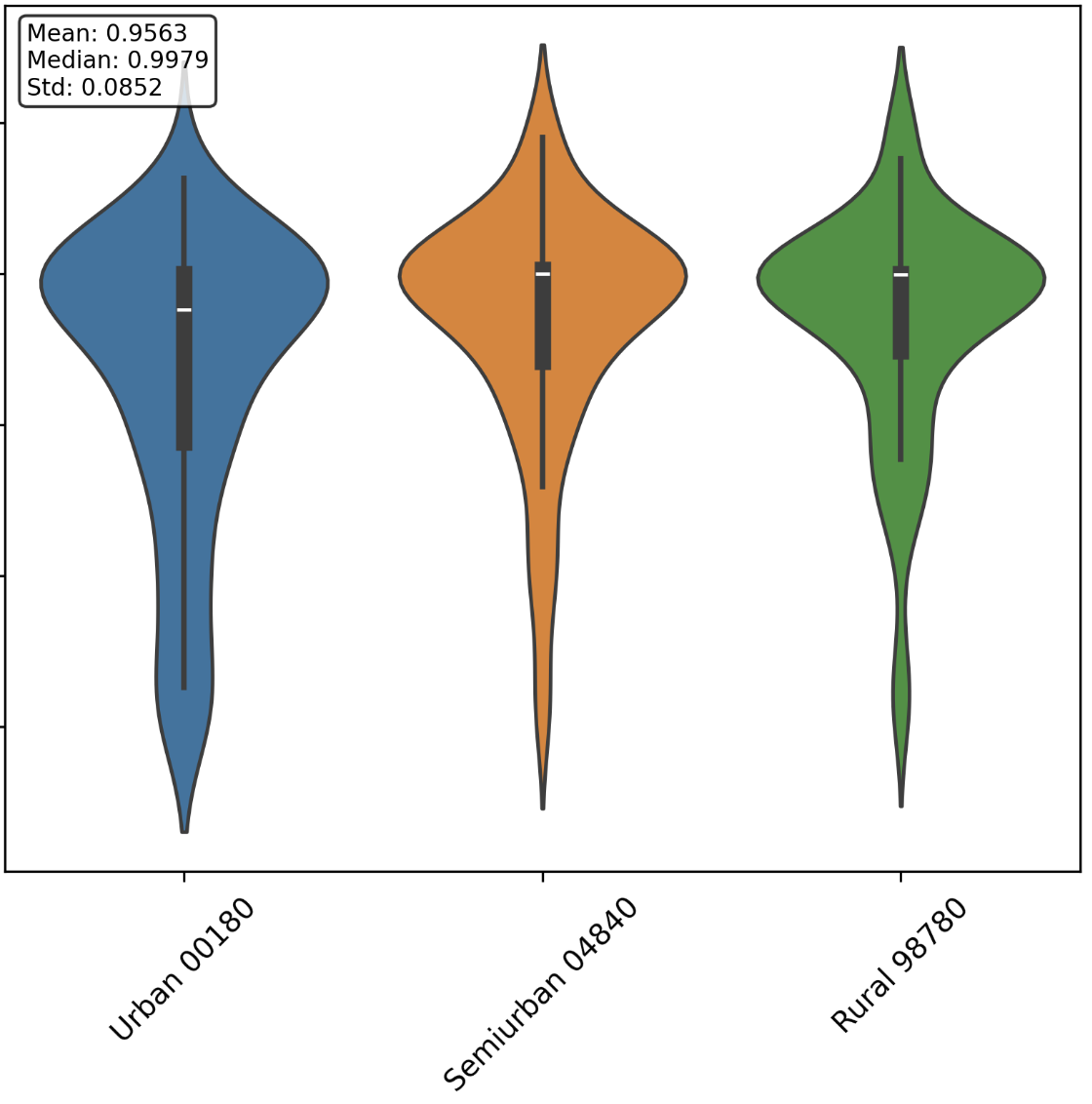}
        \caption{Relative Distortion}
        \label{fig:distortion}
    \end{subfigure}
    \caption{Violing graphs of our metrics in three different postal code areas. Top and bottom 5\% of values were capped off for clarity of the graph.}
    \label{fig:area_comparison_capped}
\end{figure}

In Figure \ref{fig:area_comparison_capped} we show a typical set of results where we have sampled 100 triangles from three different areas: 
\begin{enumerate}
    \item An urban postal code area of Kamppi-Ruoholahti (00180).
    \item A semi-urban area of Hautj\"arvi (04840).
    \item A rural area of Maaninkavaara (98780).
\end{enumerate}

\begin{figure}
    \centering
    \begin{subfigure}{0.3\textwidth}
        \centering
        \includegraphics[width=\textwidth]{img/kamppi-ruoholahti_roadmap.pdf}
        \caption{Kamppi-Ruoholahti}
    \end{subfigure}
    \hfill
    \begin{subfigure}{0.3\textwidth}
        \centering
        \includegraphics[width=\textwidth]{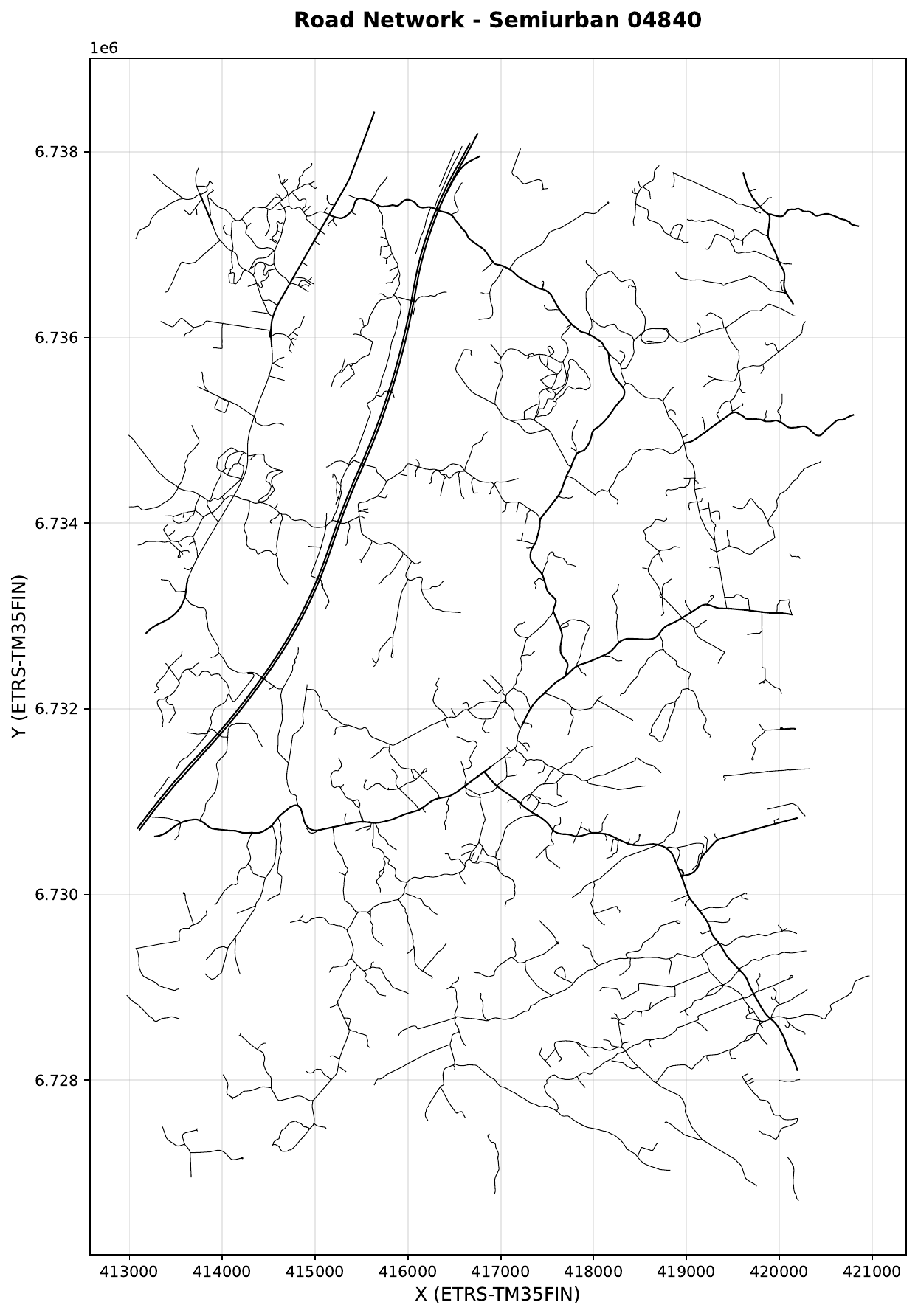}
        \caption{Hautjärvi}
    \end{subfigure}
    \hfill
    \begin{subfigure}{0.3\textwidth}
        \centering
        \includegraphics[width=\textwidth]{img/maaninkavaara_roadmap.pdf}
        \caption{Maaninkavaara}
    \end{subfigure}
    \caption{The roadmaps of our base triplet.}
    \label{fig:roadmaps_comparison}
\end{figure}

The road maps of these areas are shown in Figure \ref{fig:roadmaps_comparison}. The immediate striking fact is that the distributions of our three different metrics do seem quite dissimilar in these areas!

\begin{figure}
    \centering
    \begin{subfigure}{0.3\textwidth}
        \centering
        \adjustbox{trim=0 {.1\height} 0 0,clip}%
            {\includegraphics[width=\textwidth]{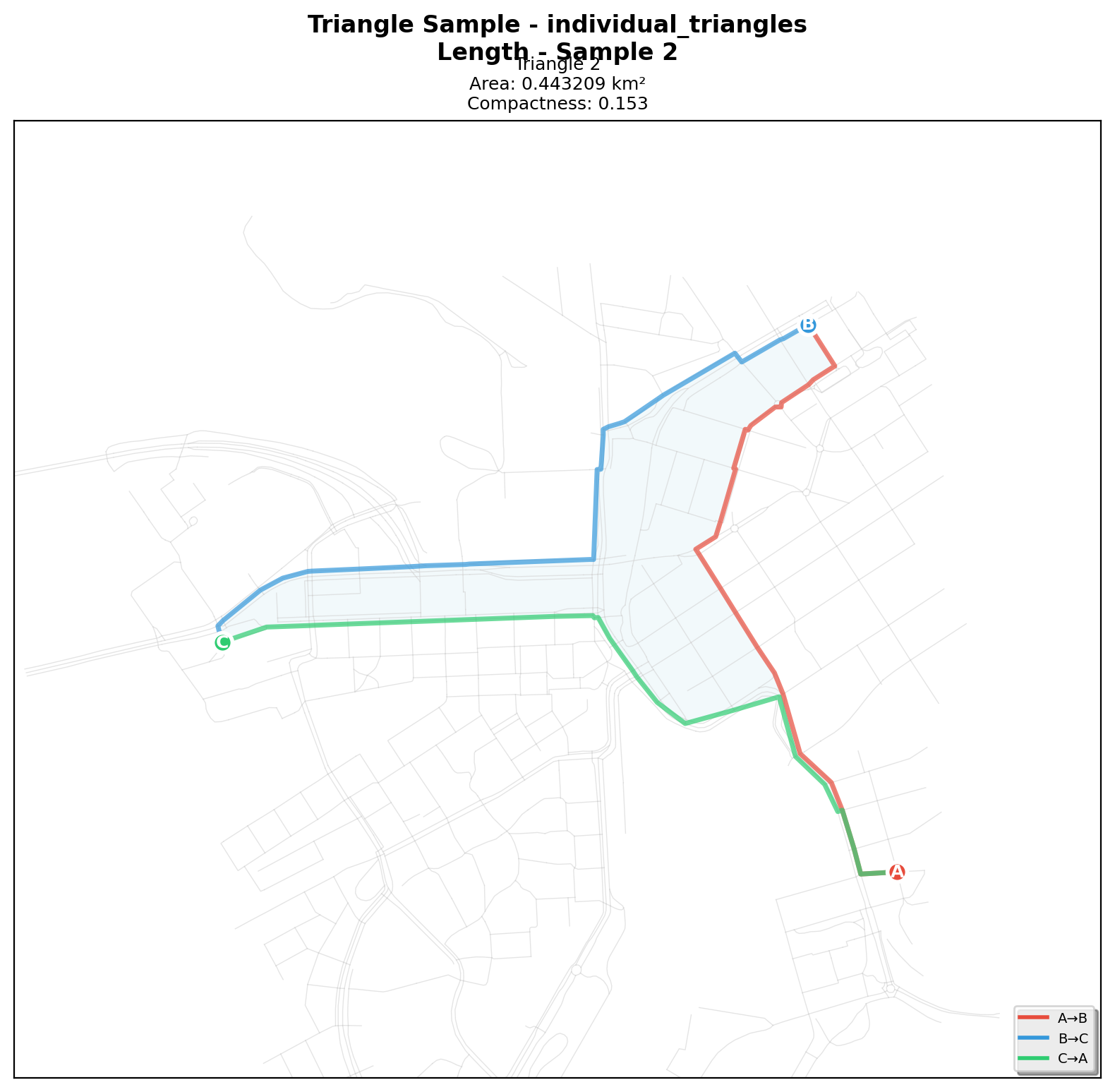}}
        \caption{Kamppi-Ruoholahti}
    \end{subfigure}
    \hfill
    \begin{subfigure}{0.3\textwidth}
        \centering
        \adjustbox{trim=0 {.1\height} 0 0,clip}%
            {\includegraphics[width=\textwidth]{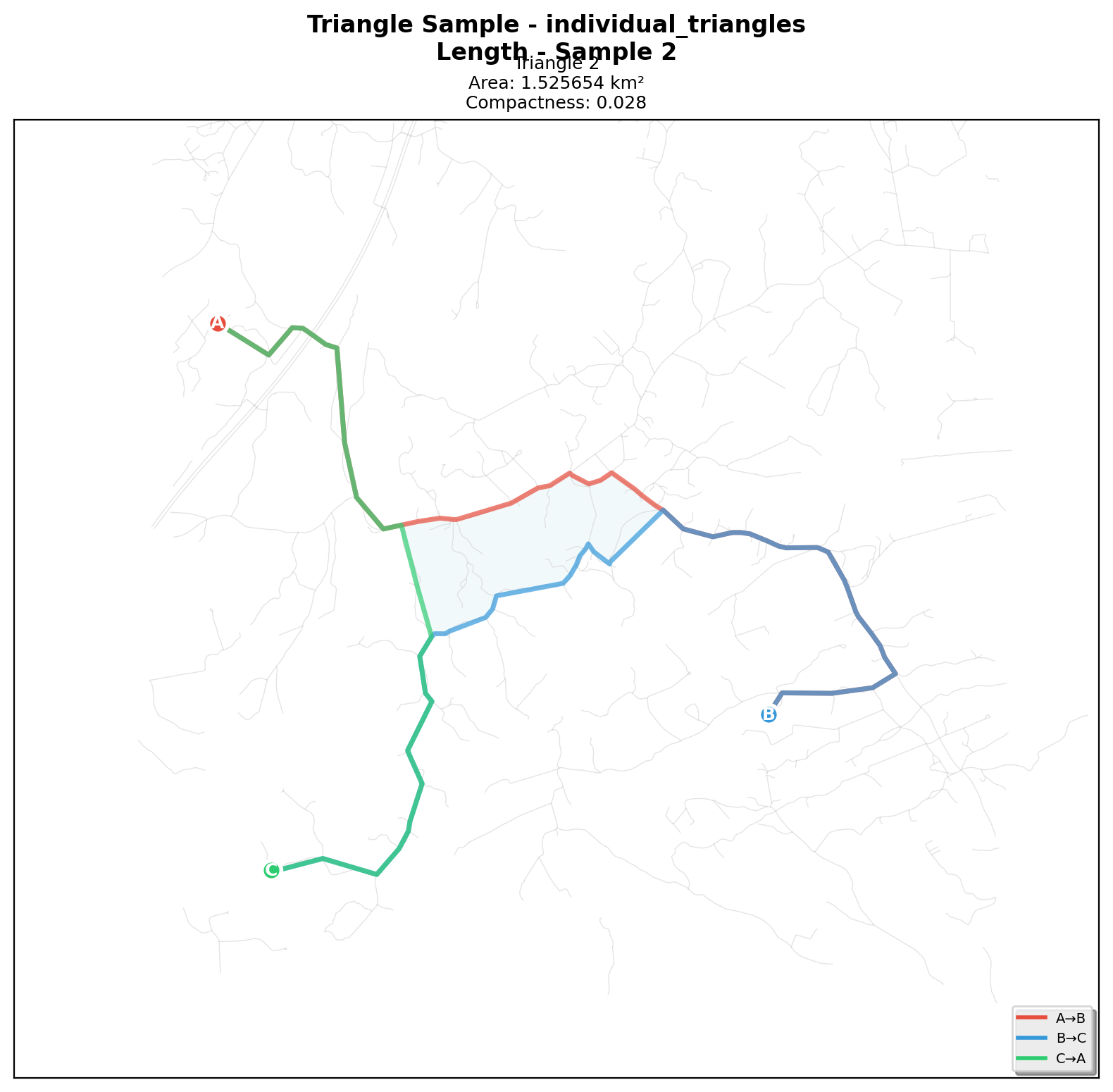}}
        \caption{Hautjärvi}
    \end{subfigure}
    \hfill
    \begin{subfigure}{0.3\textwidth}
        \centering
        \adjustbox{trim=0 {.1\height} 0 0,clip}%
            {\includegraphics[width=\textwidth]{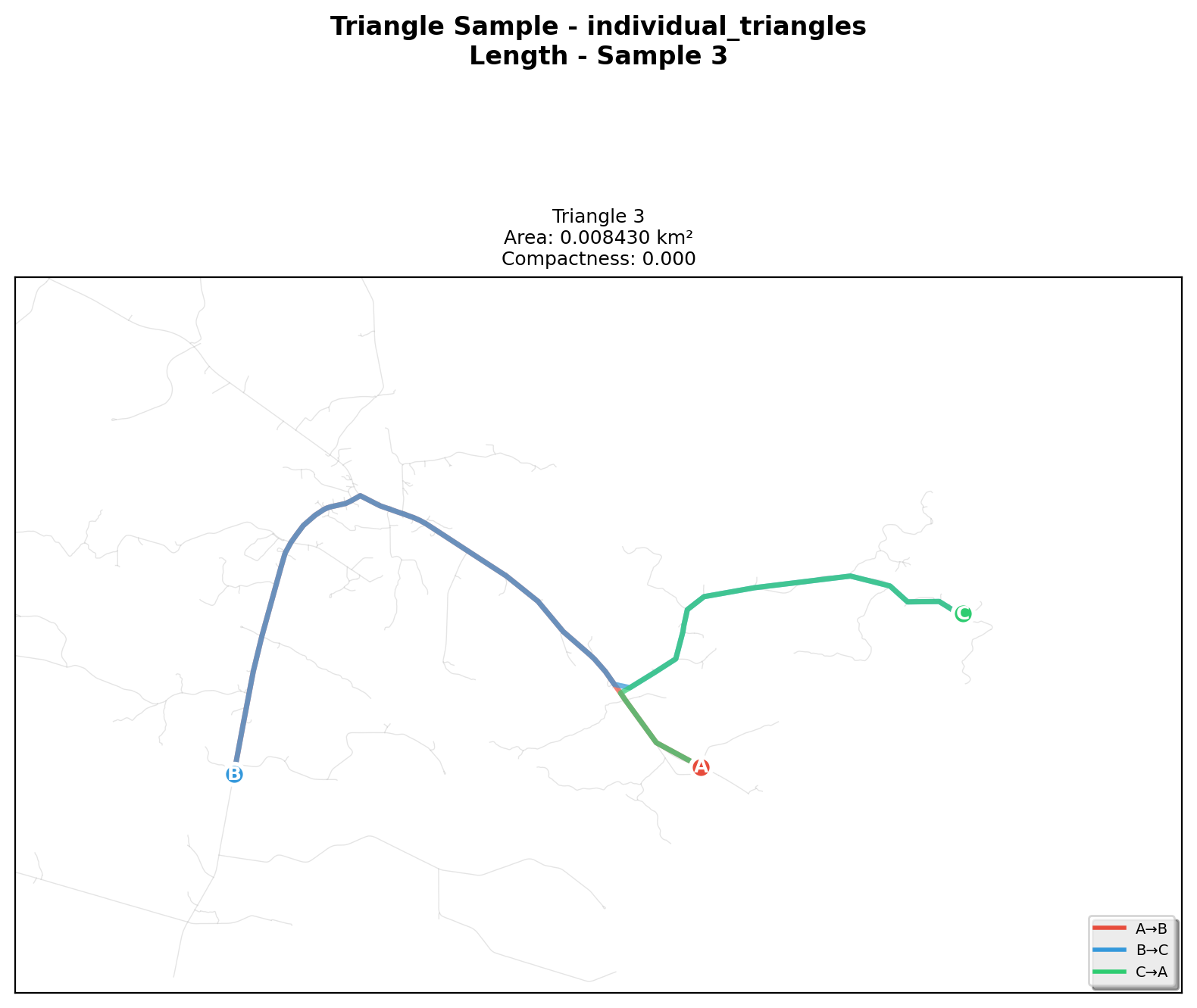}}
        \caption{Maaninkavaara}
    \end{subfigure}
    \caption{Somewhat typical sampled triangles from our triplet.}
    \label{fig:triangles_comparison}
\end{figure}

In Figure \ref{fig:triangles_comparison} we show a somewhat typical example of a sampled triangle from each of the three areas. Though we do note that even the more urban areas tend to contain some extremely hyperbolic ``degenerate'' triangles, which is part of the reason why in Figure \ref{fig:area_comparison_capped} we've capped the top and bottom $5\%$ of values in the distributions of measured values.

\subsection{Results from a larger sampling}
\label{sec:larger_sampling_results}

\begin{figure}
    \centering
    \begin{subfigure}{0.48\textwidth}
      \centering
      \adjustbox{trim={.5\width} 0 0 0,clip}
        {\includegraphics[width=2\textwidth]{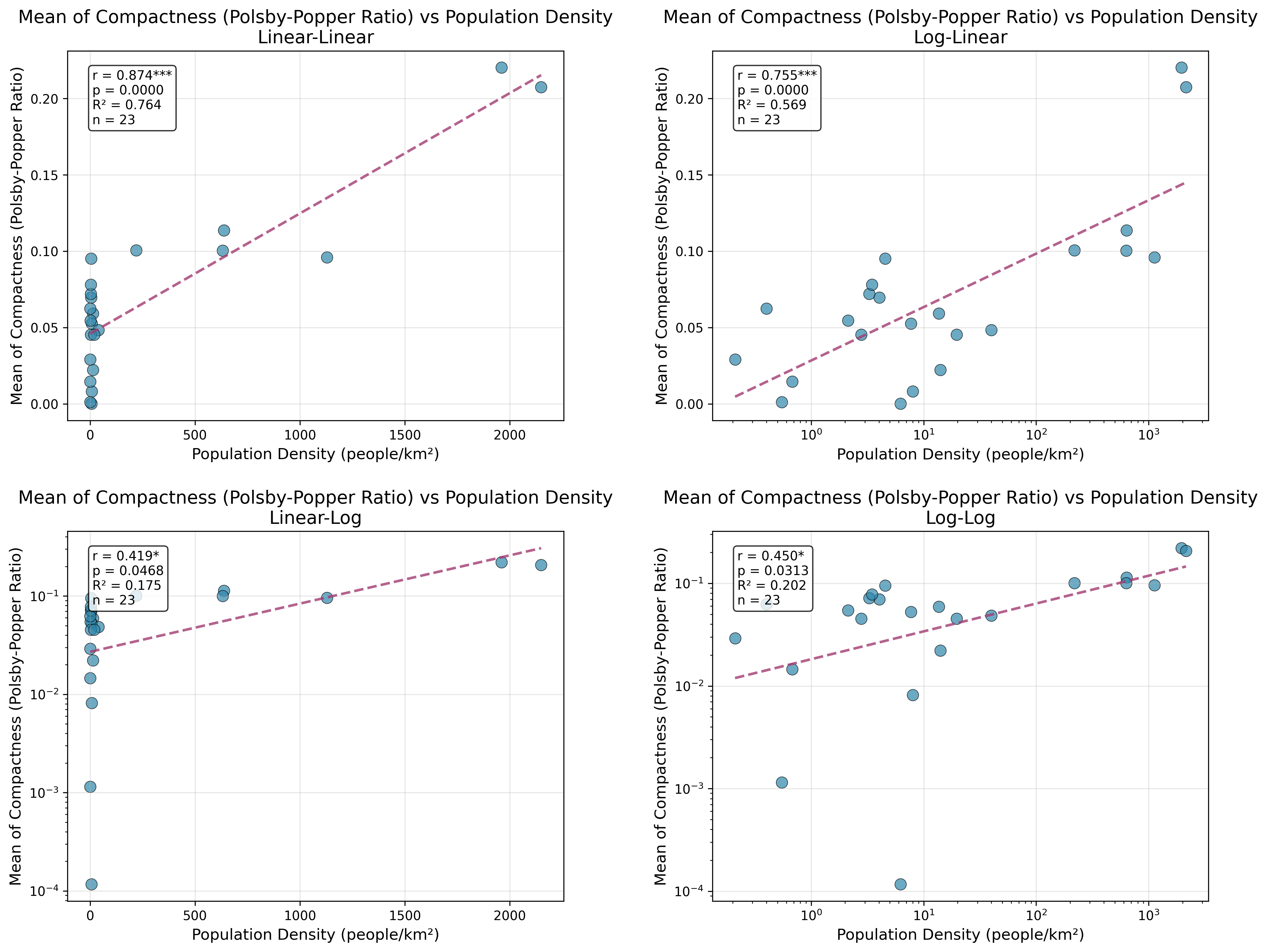}}
      \caption{Population vs mean compactness (length).}
    \end{subfigure}
    \hfill
    \begin{subfigure}{0.48\textwidth}
      \centering
      \adjustbox{trim={.5\width} 0 0 0,clip}
        {\includegraphics[width=2\textwidth]{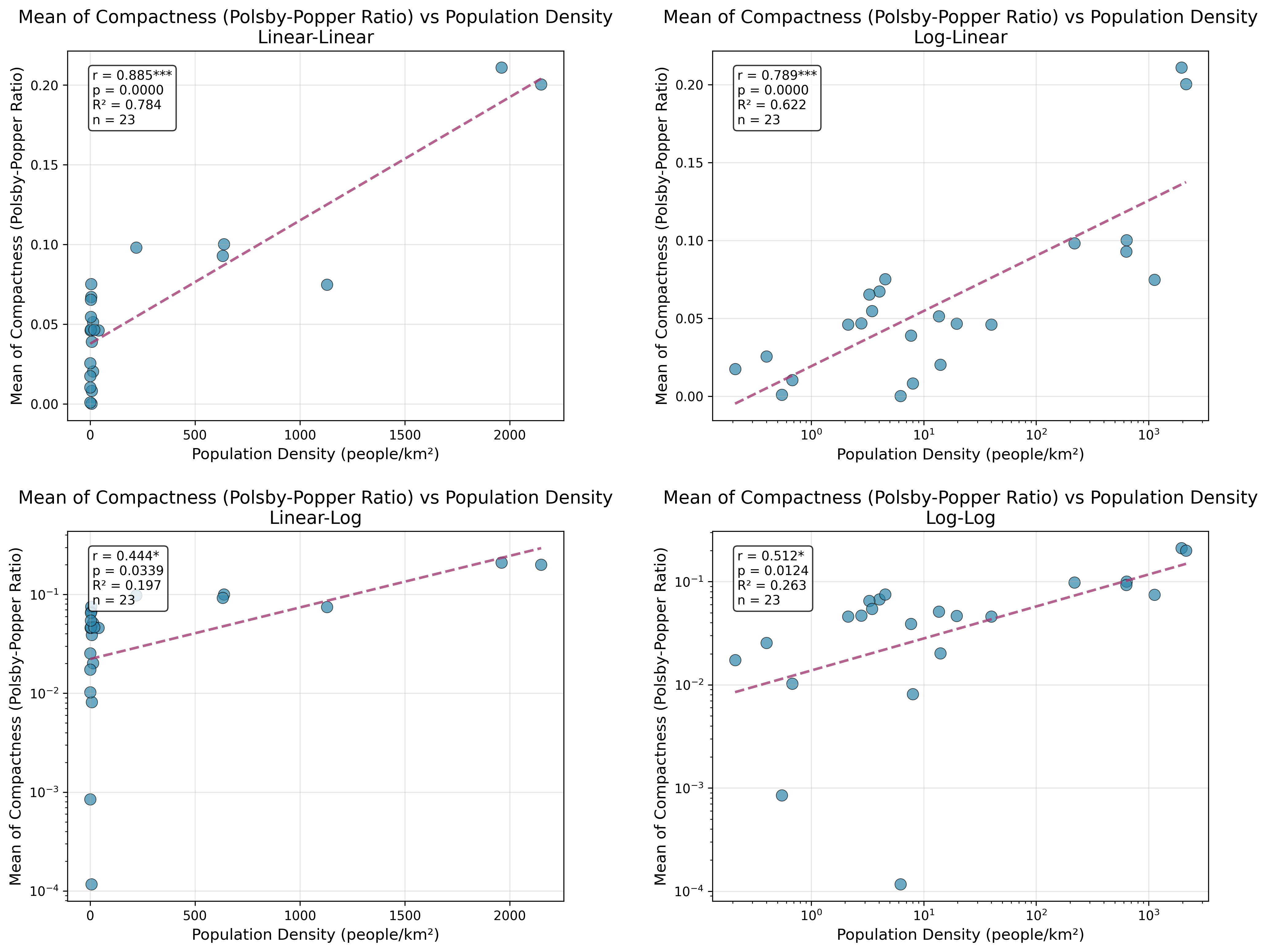}}
      \caption{Population vs mean compactness (travel time).}
    \end{subfigure}
    \caption{Correlation analysis between population density and mean compactness.}
    \label{fig:population_vs_compactness}
  \end{figure}

The Paavo dataset contained a feature column that classified postal area codes to urban, semiurban and rural areas. So we did a random sampling of 8 from each of them, and from each of the 24 areas we then sampled 100 triangles, with both length and travel time based geodesics. Motivated by Figure \ref{fig:area_comparison_capped} we tried to capture the shape of the distribution by comparing the population density of a postal code against:
\begin{itemize}
\item The mean of each of the measures.
\item The average of each of the measures.
\item The first or third quartile of each of the measures.
\item The standard deviation of each of the measures.
\item The FWHM of each of the measures.
\end{itemize}
From these the most promising candidate for visually clear correlations was the mean of the compactness measure, i.e.\ the isoperimetricity of triangles. Though we emphasize here that we \emph{did not register a prediction beforehand}, meaning that any correlations seen here might be pure chance\footnote{See e.g.\ \url{https://en.wikipedia.org/wiki/Data_dredging} or \url{https://xkcd.com/882/}.}.
In Figure \ref{fig:population_vs_compactness} we nevertheless show how the mean of the compactness measure matches against the population density, both in log-linear and log-log scales. We show the compactness measure calculated both w.r.t.\ the distance and travel time metrics of the graph.

We note that the correlation is not linear, but the data does seem to show some trends. In particular, there would be some hope in differentiating the high population density areas from less dense ones. We furthermore note that the difference between the travel time and length triangles had very little effect on the general shape of the data. (Note that in this analysis run we generated the same triplet of points for both travel time based and length based geodesic triangle approach, which explains why the data points are almost identical.)

\begin{figure}
    \centering
    \adjustbox{trim={0.5\width} {0.5\height} 0 {0.05\height},clip}
        {\includegraphics[width=2.0\textwidth]{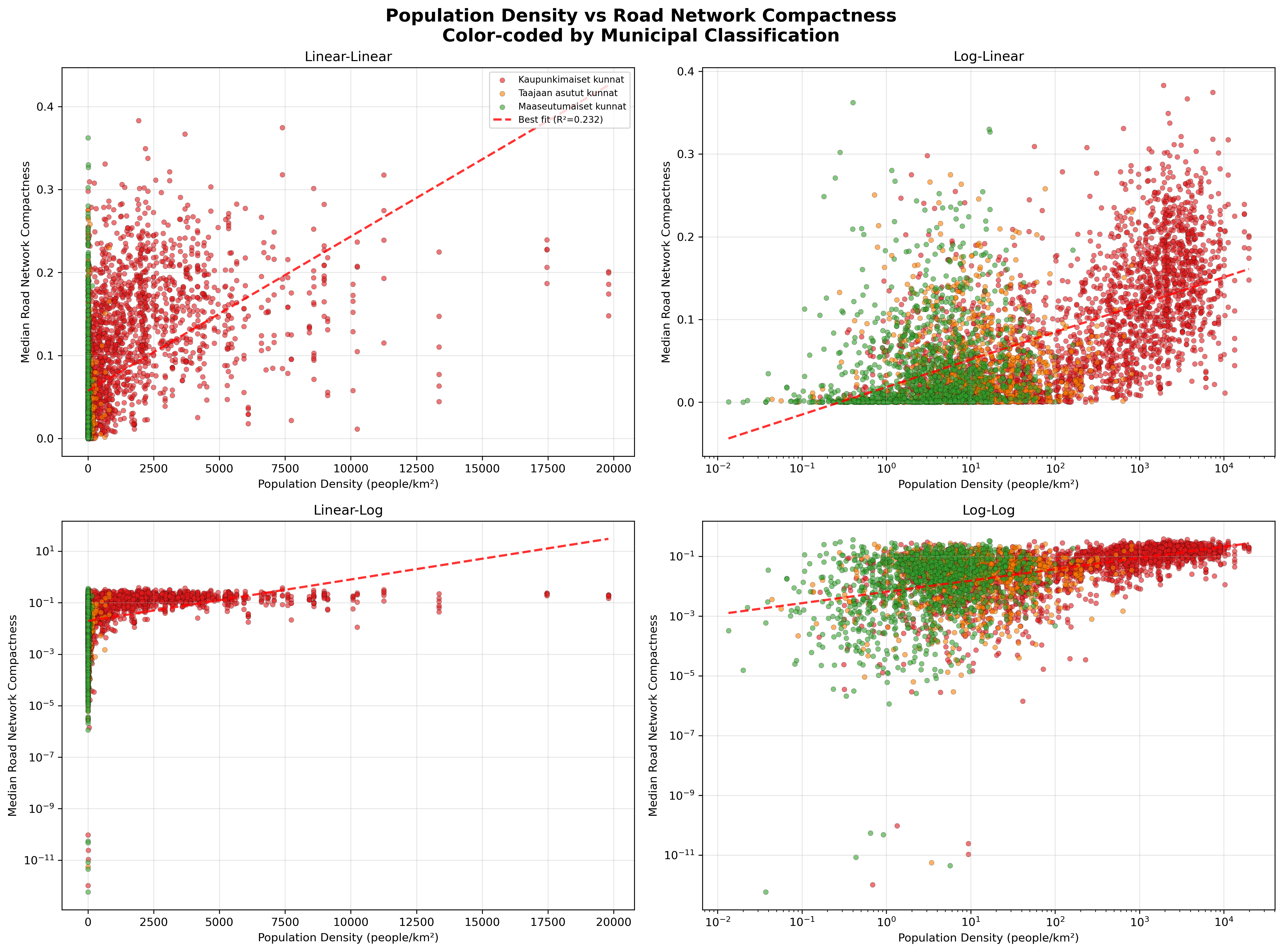}}
    \caption{Population density vs mean compactness. Using the Paavo dataset labeling, red dots are urban postal codes, orange ones semiurban and green ones rural.}
    \label{fig:large_population_vs_compactness_analysis}
\end{figure}


\section{Final thoughts}
\label{sec:conclusions}

For us the violin graphs like the one in Figure \ref{fig:area_comparison_capped} feel like strong evidence that there indeed is something about the rurality of road networks that corresponds to some measures of hyperbolicity. The results of Section \ref{sec:larger_sampling_results} encourage this idea, though from them it is obvious that the connection is not trivial. More nuanced statistical measures are probably needed. Furthermore, we think that the division of areas based on postal codes is quite artificial and produces artefacts to our measurements.

\begin{figure}
    \centering
    \includegraphics[width=0.8\textwidth]{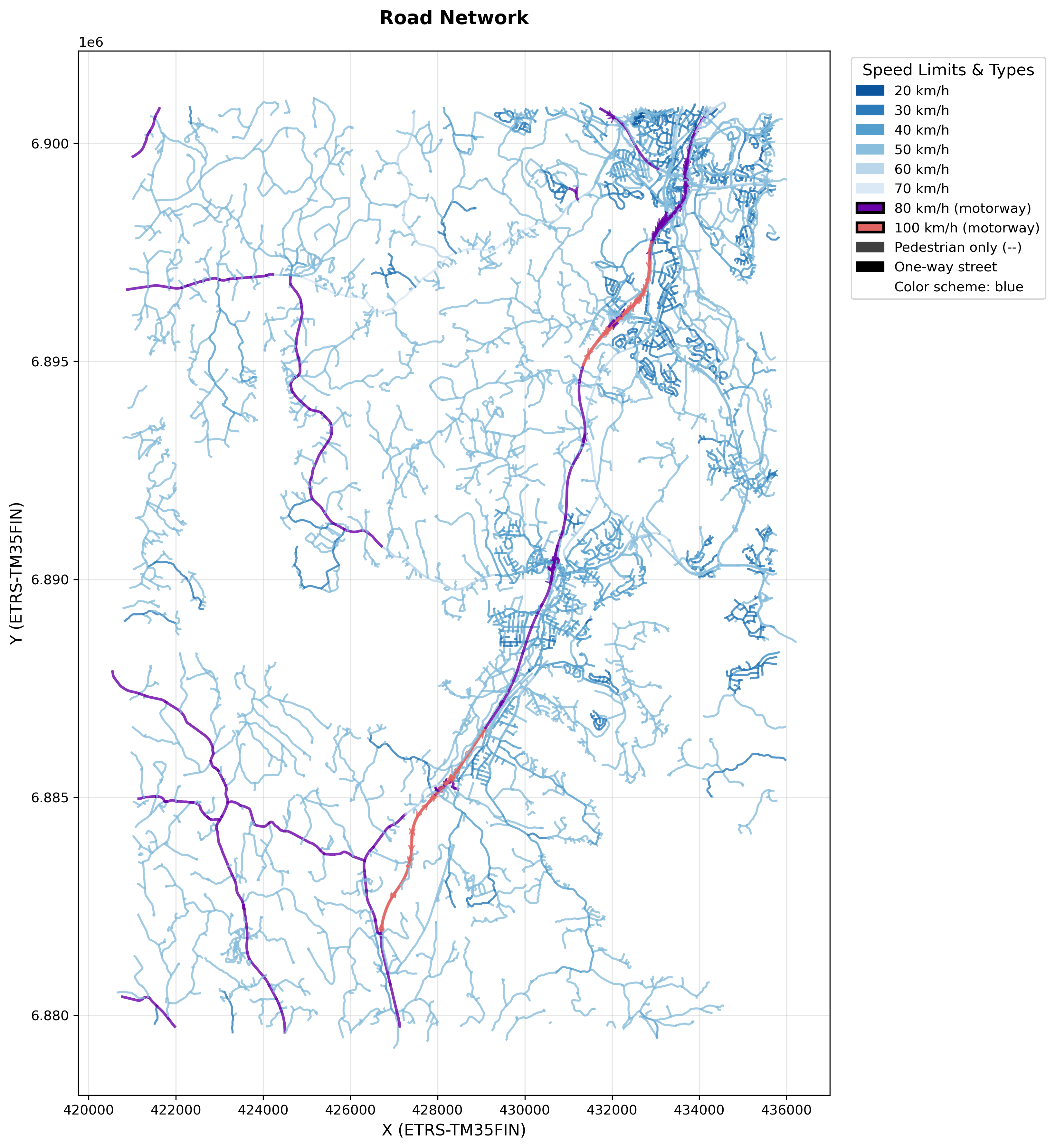}
    \caption{The Muurame area road network.}
    \label{fig:muurame_road_network}
\end{figure}

As a simple example of the artificiality of postal codes, in Figure \ref{fig:muurame_road_network} we note that a given postal code area is not very homogenous. It has some high-density areas, some more hyperbolic-looking rural areas and also a narrow ``funnel'' in the center. This will have a big effect on the hyperbolicity measures, as any triplet of points with points of either side of this funnel will have to squeeze itself through this narrow area, making its geometry more hyperbolic in the eyes of our measures. 

\begin{figure}[htbp]
    \centering
    \begin{subfigure}{0.45\textwidth}
        \centering
        \includegraphics[width=\textwidth]{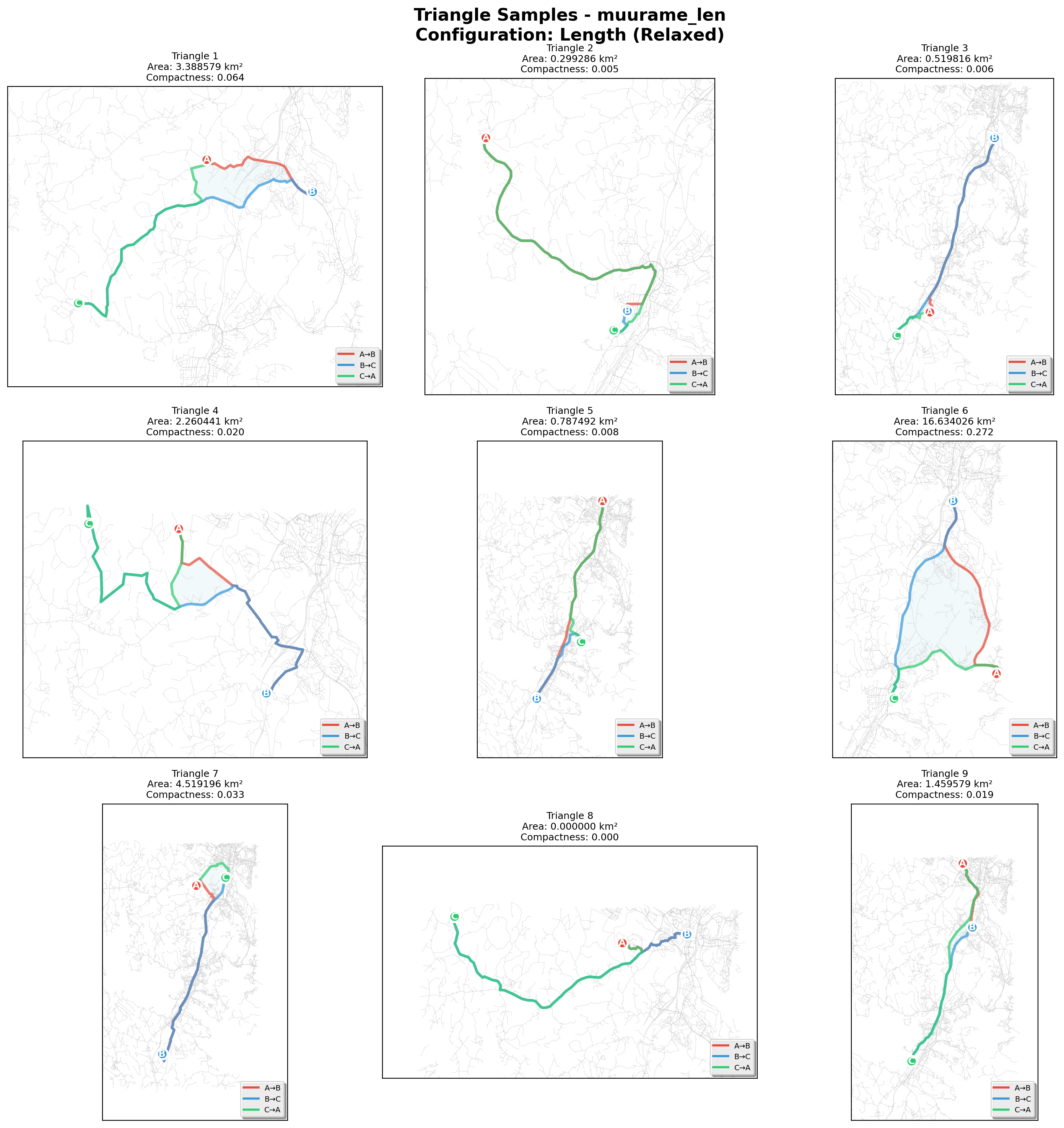}
        \caption{Muurame: triangles by length metric.}
    \end{subfigure}
    \hfill
    \begin{subfigure}{0.45\textwidth}
        \centering
        \includegraphics[width=\textwidth]{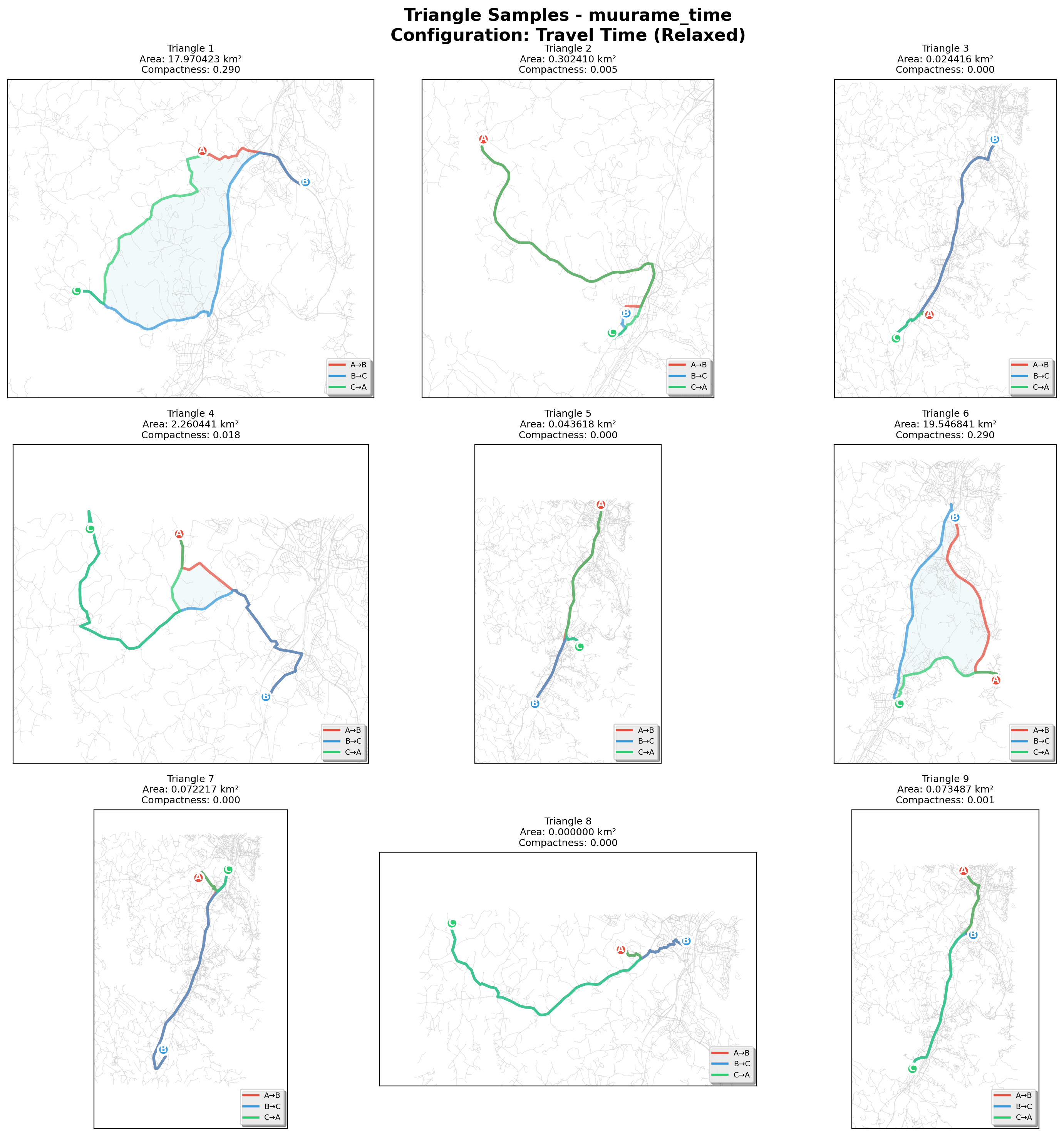}
        \caption{Muurame: triangles by travel time metric.}
    \end{subfigure}
    \caption{Examples of triangles sampled in the Muurame area, visualized using both the length and travel time metrics.}
    \label{fig:muurame_len_time_triangles}
\end{figure}

Furthermore we note that the existence of a motorway going through the area can strong effects on the travel-time geometry of road network. Though in the particular case of Muurame, we see in Figure \ref{fig:muurame_len_time_triangles} that for 9 randomly sampled point triplets the geometry is not vastly different for the triangles. Numerical results are maybe slightly more hyperbolic for the travel time metric, but not much. We conjecture that in many cases the fast roads tend also to be more straight, and there arent massive differences in the resulting geometry, at least on the scale of postal code area distances.

\subsection{Further possible avenues of research}
\label{sec:furhter_ideas}

We hesitate to call these further \emph{research} problems, but there are several ideas that we did not have time to explore here, and doubt we have time to dedicate in the near future. We list them here as possible project ideas for anyone wishing to explore the ideas here further. We again remind that the code for our analysis tool is freely available at \url{https://github.com/ramiluisto/RoadGeometry}.

\begin{enumerate}
    \item Analyzing different levels of domains. We've limited ourselves to postal codes. It would be nice to look at some data from larger areas as well, especially after optimizing the code a bit. 

    \item Are there other methods we could better apply here for hyperbolicity measures? E.g.\ the travel speed approach briefly mentioned in Section \ref{sec:GeneralHyperbolicity} is mostly untouched upon, and the existing methods could be fine-tuned. Also the hausdorff distance approach could be done with graph-internal geodesics if one optimized the path-finding part a bit. In all of the approaches, our statistical analyses were quite naive and would probably benefit from deeper thought.

    \item Another very different approach to analyzing hyperbolicity would be to look at the amount of ``almost geodesic'' paths between a pair of points. Once could e.g.\ sample pairs of points $a,b$, find a geodesic $\alpha \colon a \curvearrowright b$ connecting them and then see how many paths $\beta \colon a \curvearrowright b$ there exists with $\ell(\beta) \leq \ell(\alpha) + \eps$ for various values of $\eps$. In a highly hyperbolic space geodesics are sort of ``fragile'' and we predict that more hyperbolic areas see less ``almost geodesic'' (or \emph{quasigeodesic}) paths than more flat areas. This would most likely require some preprocessing of the graph data to be computationally feasible.

    \item The issues discussed in relation to Figure \ref{fig:muurame_road_network} bring to mind some ideas related to the Poincare inequality or conformal modulus -- is there something to look at there?

    \item Having played various computer games\footnote{E.g.\ Mini Motorways or Cities Skylines.} that simulate some aspects of city building, we note that some of the structures here might be emergent properties of some heuristical energy minimization? Is there a difference in ``evolved'' vs.\ ``planned'' road systems? Could we create simulations on how populations might want to create road connections and observe similar statistics of hyperbolicity as in here? See also Geoffrey B.\ West, \emph{Scale: The Universal Laws of Growth, Innovation, Sustainability, and the Pace of Life in Organisms, Cities, Economies, and Companies}.

    \item The sampling approach we use is very naive, and we often sampled point triplets almost from the same line. It might be better to try to sample triangles in a way that biases the samples towards more equilateral triangles.
    

\end{enumerate}

\bibliographystyle{alpha}
\bibliography{bibliography}

\end{document}